\documentclass[prd,aps,twocolumn]{revtex4}
\usepackage{amssymb,amsmath,amsfonts,amsbsy,epsfig}

\newcommand{\be}{\begin{equation}}
\newcommand{\ee}{\end{equation}}
\newcommand{\bea}{\begin{eqnarray}}
\newcommand{\eea}{\end{eqnarray}}
\newcommand{\nn}{\nonumber}

\def\k{{\bf k}}

\def\x{{\bf x}}
\def\y{{\bf y}}

\def\R{{\mathcal R}}
\def\P{{\mathcal P}}
\def\C{{\mathcal C}}
\def\O{{\mathcal O}}

\begin{document}

\title{B-modes and the sound speed of primordial fluctuations}
\date{\today}

\author{
Gonzalo A. Palma$^{a}$ and Alex Soto$^{b}$
}

\affiliation{
$^{a}$Departamento de F\'{i}sica, Facultad de Ciencias F\'{i}sicas y Matem\'{a}ticas, Universidad de Chile, \mbox{Blanco Encalada 2008, Santiago, Chile} \\
$^{b}$Departamento de F\'{i}sica, Facultad de Ciencias, Universidad de Chile, \mbox{Las Palmeras 3425, \~Nu\~noa, Santiago, Chile}
}

\begin{abstract} 

It was recently shown that a large value of the tensor to scalar ratio $r$ implies a constraint on the minimum value of the sound speed $c_s$ of primordial curvature perturbations during inflation that is stronger than current bounds coming from non-Gaussianity measurements. Here we consider additional aspects related to the measurement of B-modes that may provide additional leverage to constrain the sound speed parametrizing non-canonical models of inflation. We find that a confirmation of the consistency relation $r = - 8 n_t$ between the tensor to scalar ratio $r$ and the tensor spectral index $n_t$ is not enough to rule out non-canonical models of inflation with a sound speed $c_s$ different from unity. To determine whether inflation was canonical or not, one requires knowledge of additional parameters, such as the running of the spectral index of scalar perturbations $\alpha$. We also study how other parameters related to the ultra violet completion of inflation modify the dependence of $r$ on $c_s$. For instance, we find that heavy degrees of freedom interacting with curvature fluctuations generically tend to make the constraint on the sound speed stronger. Our results, combined with future observations of primordial B-modes, may help to constrain the background evolution of non-canonical models of inflation.

\end{abstract}

\maketitle 

\section{Introduction}

The recent claim that the tensor to scalar ratio $r$ might be of order $\sim 0.1$, as implied by BICEP2 detection of primordial B-modes~\cite{Ade:2014xna}, has reinvigorated the debate about the fundamental nature of inflation~\cite{Guth:1980zm, Linde:1981mu, Albrecht:1982wi, Starobinsky:1980te, Mukhanov:1981xt} among early universe cosmologists. Although BICEP2 results have been subject to revision~\cite{Flauger:2014qra, Mortonson:2014bja, Adam:2014bub, Kamionkowski:2014wza, Ade:2015tva}, the prospects of living in a universe that owes its structure to a large-scale inflationary phase has driven us to clarify the role of inflation as the mechanism to generate primordial fluctuations~\cite{Cook:2014dga, Martin:2014lra, Abazajian:2014tqa, Ashoorioon:2014nta}. In ref.~\cite{Kinney:2014jya}, for instance, it was argued that a large value of $r$ favors a substantially large and negative running of the spectral index, necessary to accommodate the upper bound on $r$ inferred from Planck data~\cite{Ade:2013uln, Smith:2014kka}. This in turn, would prevent quantum fluctuations dominate over the classical evolution of the inflaton, precluding the possibility of having eternal inflation~\cite{Linde:1986fc}. 

Another important consequence of having a large tensor to scalar ratio is that the inflaton must have had super-Planckian excursions in field space~\cite{Lyth:1996im, Baumann:2011ws}. For this to be possible without fine tuning, one requires the presence of a (shift) symmetry at the effective single field theory level, descending from an ultra violet (UV) complete theory, presumably involving the existence of several degrees of freedom gravitationally coupled together. Building a satisfactory UV theory embedding inflation, in a well established framework such as supergravity and/or string theory ---incorporating a shift symmetry--- constitutes one of the long standing challenges within the study of early universe cosmology~\cite{Kachru:2003sx, Conlon:2005jm, Silverstein:2008sg, Flauger:2008ad, Baumann:2011nk, Baumann:2014nda, Chialva:2014paa}. A large value of $r$ (say, in the range $0.01$-$0.1$) would rule out a large class of models, and would provide an important hint toward the mass scale characterizing the geometry of fundamental theories incorporating inflation~\cite{McAllister:2008hb, Covi:2008ea, Covi:2008cn, Kallosh:2010xz, Borghese:2012yu, Roest:2013aoa}. In addition, it could give us important information about the structure of couplings between the inflaton field and the standard model~\cite{Burgess:2004kv, Hardeman:2010fh}.

\begin{figure}[t!]
\includegraphics[scale=0.4]{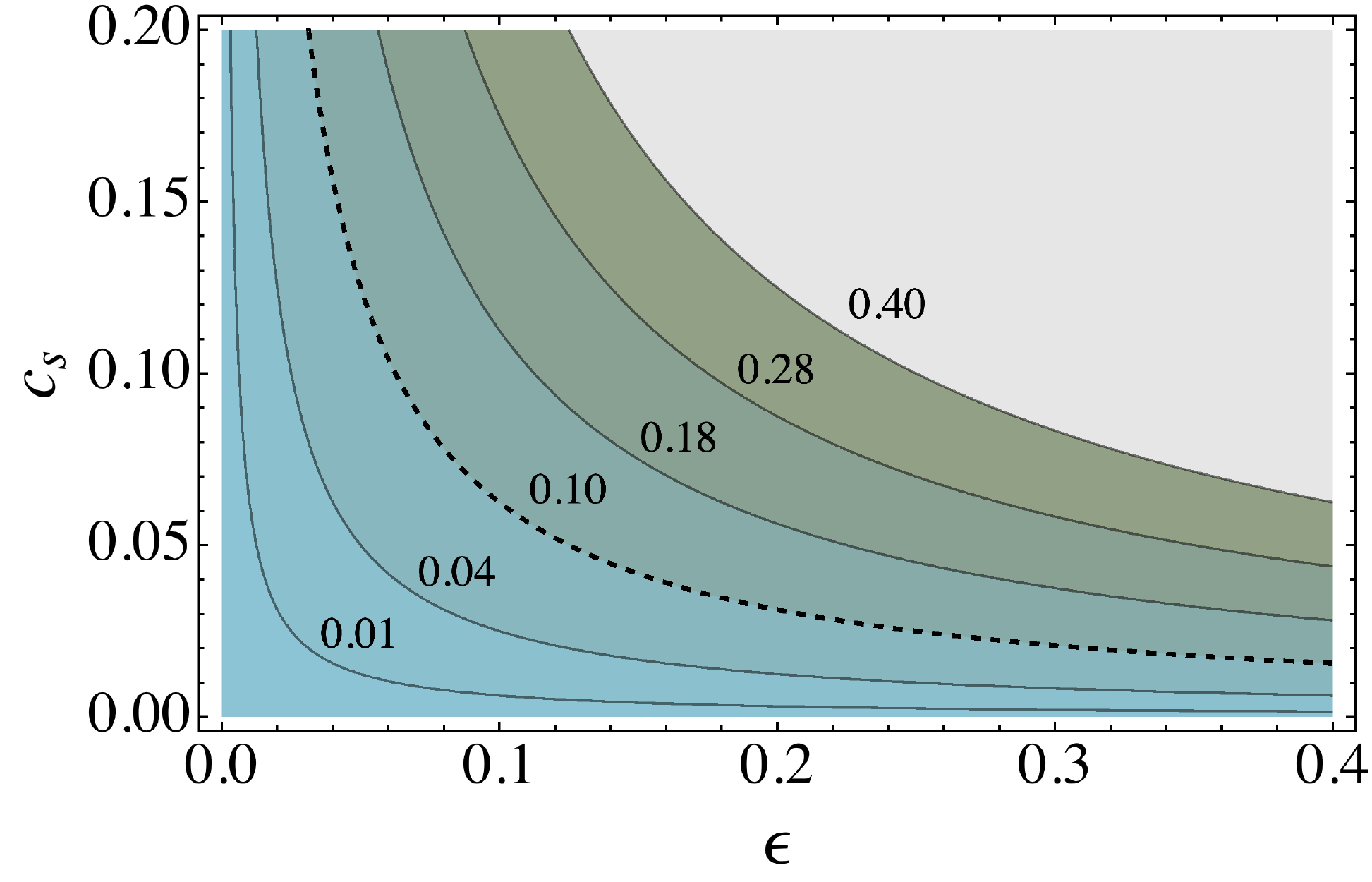}
\caption{The figure shows the contour plot for $r$ in the $\epsilon$-$c_s$ plane, obtained from eq.~(\ref{basic-r}). The dashed line corresponds to the case for $r = 0.1$.}
\label{fig:ec-basic}
\end{figure}

\begin{figure}[t!]
\includegraphics[scale=0.4]{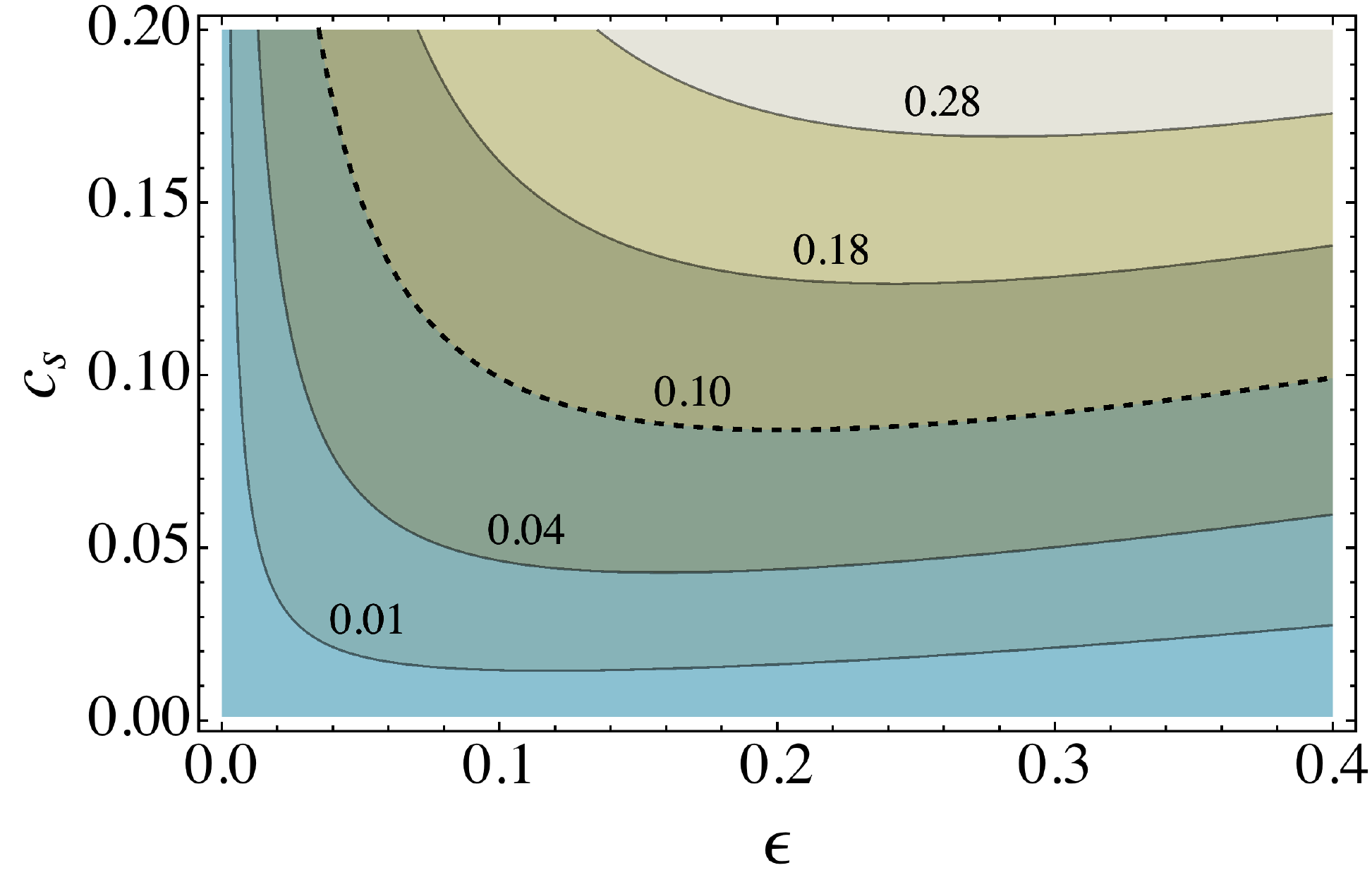}
\caption{The figure shows the contour plot for $r$ in the $\epsilon$-$c_s$ plane, obtained from eq.~(\ref{basic-r-corr}). The dashed line shows the case $r = 0.1$, which implies a lower bound $c_s > 0.09$.}
\label{fig:ec-basic-log}
\end{figure}

Interestingly, a large value of $r$ opens up the possibility of placing new constraints on parameters that were previously thought to remain degenerate at the level of two-point correlation functions measured in Cosmic Microwave Background (CMB) experiments. Indeed, it is well known that, to lowest order in slow-roll, the tensor to scalar ratio is determined by a combination of the slow-roll parameter $\epsilon$ [defined in eq.~(\ref{def-epsilon})] and the sound speed $c_s$ of adiabatic perturbations, as:
\be
r = 16 \epsilon c_s . \label{basic-r}
\ee
Depending on the measured value of $r$, eq.~(\ref{basic-r}) gives us different possible values for $\epsilon$ and $c_s$ as shown in FIG.~\ref{fig:ec-basic}.
The degeneracy between $\epsilon$ and $c_s$ implied by eq.~(\ref{basic-r}) may be resolved with the help of measurements of non-Gaussianity~\cite{Chen:2006nt, Senatore:2009gt}, which currently imply $c_s > 0.02$~\cite{Ade:2013ydc, Ade:2015ava}. However, the authors of ref.~\cite{Baumann:2014cja} found that a large value of $r$ grants the possibility of placing a stronger lower bound on $c_s$ (see also~\cite{Zavala:2014bda} for another analysis on the degeneracy implied by $c_s$). The crucial observation leading to this result is that the lapse of time between the horizon crossings of scalar and tensor modes implied by $c_s \neq 1$ comes together with a sizable running of the Hubble parameter $H$, modifying the way that $r$ depends on both $c_s$ and $\epsilon$, giving us back~\cite{Agarwal:2008ah, Powell:2008bi}:
\be
r = 16 \epsilon c_s e^{2 \epsilon \ln c_s} . \label{basic-r-corr}
\ee
Here, the quantity $2 \epsilon \ln c_s$ is due to the running of the Hubble parameter $H$ between the two horizon crossing times. Because $2 \epsilon \ln c_s$ can take values of order $1$ without violating the slow-roll condition $\epsilon \ll 1$, one deduces from (\ref{basic-r-corr}) that $c_s$ is bounded from below, as shown in FIG.~\ref{fig:ec-basic-log}.
In particular, (\ref{basic-r-corr}) implies that $c_s > 0.14$ for the range of values $r > 1.3$. Moreover, one sees that bounds on $c_s$ are stronger than current non-Gaussian constraints as long as $r > 0.01$ \footnote{As shown in ref.~\cite{D'Amico:2014cya}, it is still possible to reconcile large non-Gaussianity with a large value of the tensor to scalar ratio $r$. This is because in the effective field theory of inflation there is a second parameter ---in addition to the sound speed--- capable to enhance deviations from Gaussianity, but with a very constrained shape.}.

Models with $c_s \neq 1$ appear in a number of different contexts where inflation is driven by a nontrivial fluid~\cite{ArmendarizPicon:1999rj} as typically found in low energy compactifications of string theory~\cite{Silverstein:2003hf, Alishahiha:2004eh}. For example, in multi field models, curvature perturbations are forced to propagate with a sound speed $c_s < 1$ whenever heavy fields interact with the inflaton as a result of turns of the inflationary trajectory in field space~\cite{Tolley:2009fg, Cremonini:2010ua, Achucarro:2010da, Shiu:2011qw, Cespedes:2012hu, Achucarro:2012sm, Avgoustidis:2012yc, Burgess:2012dz, Gao:2013ota}. More generally, the sound speed plays a central role in the effective field theory approach to inflation~\cite{Cheung:2007st, Senatore:2010wk} which provides a systematic parametrization of departures from the standard single field canonical case. 

The mere fact that $c_s$ could be different from $1$ requires physics beyond the canonical single field paradigm in which slow-roll inflation is based. It is therefore sensible to expect additional parameters related to the physics underlying $c_s \neq 1$, due to operators that belong to the UV completion of the effective theory describing inflation~\cite{Baumann:2011su}. For instance, in the case of heavy fields interacting with the inflaton, one expects a non vanishing running of the sound speed $s = \dot c_s / H c_s$~\cite{Achucarro:2012yr} together with additional nontrivial operators affecting the dynamics of the inflaton that are turned on when $c_s \neq 1$~\cite{Gwyn:2012mw, Cespedes:2013rda, Gwyn:2014doa}.

The aim of this work is to study additional effects on eq.~(\ref{basic-r-corr}) to those considered in ref.~\cite{Baumann:2014cja} implied by the running of slow-roll parameters. In particular, we pay special attention on the effects on (\ref{basic-r-corr}) coming from the running of $\epsilon$ during the lapse of time between the horizon crossing of scalar and tensor modes. We find that the dependence of $r$ on the running of $\epsilon$ is such that a confirmation of the consistency relation $r = - 8 n_t$ between the tensor to scalar ratio $r$ and the tensor spectral index $n_t$ is not enough to rule out non-canonical models of inflation with a sound speed $c_s \neq 1$. As a consequence, to determine whether inflation was canonical or not with the help of B-mode measurements, one requires knowledge of additional parameters, such as the running of the spectral index of scalar perturbations $\alpha$. 

In addition, we study the effects on (\ref{basic-r-corr}) implied by operators that take a role in the UV completion of inflationary theories with nontrivial sound speeds, which is motivated by the well known example of models where heavy fields interact with the inflaton, implying a variety of additional effects that appear along with a sound speed different from unity~\cite{Gwyn:2012mw, Cespedes:2013rda, Gwyn:2014doa}. To this extent, we adopt the effective field theory of inflation perspective~\cite{Cheung:2007st} and study the consequences of UV physics on CMB observables by taking into account operators in the action for curvature perturbation that are turned on when $c_s \neq1 $. We find that these operators modify substantially the inference of a lower bound on the sound speed, in such a way that heavy degrees of freedom interacting with the inflaton generically tend to make the constraint on the sound speed stronger.

This article is organized as follows: In Section~\ref{sec:tensor-scalar} we review the dependence of the power spectra of scalar and tensor modes on different inflationary parameters. Then, in Section~\ref{sec:constraints} we discuss the constraints on the speed of sound inferred by the results obtained in Section~\ref{sec:tensor-scalar}.  In Section~\ref{sec:uv-par} we discuss the effects of UV-operators affecting the evolution of fluctuations on the constraints on $c_s$. Finally, in Section~\ref{sec:conclusions} we discuss our results and provide some concluding remarks. We have left the more technical discussion for the appendices, where we deduce various of the expressions dealt with in the body of our work.

\section{The tensor to scalar ratio} \label{sec:tensor-scalar}

Single field slow-roll inflation is realized by a universe in a quasi-de Sitter phase characterized by a steady evolution of the Hubble parameter $H \equiv \dot a/a$ (where $a$ is the scale factor). To ensure this, one demands:
\be
\epsilon \ll 1 , \qquad \epsilon \equiv - \frac{\dot H}{H^2} . \label{def-epsilon}
\ee
The quantity $\epsilon$ is the usual first slow-roll parameter, and to parametrize its time-evolution it is customary to define a second slow-roll parameter $\eta$ as:
\be
\eta \equiv \frac{\dot \epsilon}{H \epsilon} . \label{def-eta}
\ee
To ensure that $\epsilon \ll 1$ for a sufficiently long time, one demands the additional condition $|\eta| \ll 1$. The sound speed $c_s$ at which adiabatic perturbations propagate also plays a role in determining the value of inflationary observables. To parametrize its time dependence it is useful to define 
\be
s = \frac{\dot c_s}{H c_s} ,  \label{def-s}
\ee
and ask $|s| \ll 1$ to stay consistent with the requirements of slow-roll evolution of the background~\footnote{It has been shown, however, that heavy fields may induce rapid variations of the sound speed, violating the condition $|s| \ll 1$ without necessarily implying a violation of $\epsilon \ll 1$~\cite{Achucarro:2010da, Cespedes:2012hu}.}. The amount of information stored in eqs.~(\ref{def-epsilon}), (\ref{def-eta}), and~(\ref{def-s}) allows us to derive the power spectra for scalar and tensor perturbations. In Appendix~\ref{app-pert} we review the standard perturbation theory used to study the dynamics of fluctuations at leading order in terms of the slow-roll parameters, and derive the power spectra for scalar and tensor modes. In what follows we quote the necessary results from these appendices to discuss the computation of the tensor to scalar ratio $r$.

\subsection{Computation of $r$}

Assuming that the universe was driven by a single fluid, it is straightforward to derive that the scalar power spectrum, to first order in the slow-roll parameters, is given by~\cite{Stewart:1993bc}
\bea
\P_\R (k)  &=&  \frac{H_0^2 }{8  \pi^2 \epsilon_0 c_0} \bigg[ 1 - (2 \epsilon_0 + \eta_0 + s_0) \ln \left( \frac{k c_0}{a_0 H_0} \right)  \nn \\
&& +(2 \epsilon_0 + \eta_0 + s_0) \C  - 2 (\epsilon_0 + s_0)\bigg] , \label{P-R-first-order}
\eea
where $\C \equiv 2 - \ln 2 - \gamma \simeq 0.73$ ($\gamma$ being the Euler-Mascheroni constant). This expression asumes units such that $m_{\rm Pl}^2=1$, and is derived in Appendix~\ref{app-pert}. The zeroth order value of the power spectrum is scale independent, and is given by $H_0^2 / 8  \pi^2 \epsilon_0 c_0$. The corrections inside the square brackets are due to departures from a de Sitter space-time given in terms of slow-roll parameters, and imply a scale dependent piece proportional to $\ln (k c_0 / a_0 H_0)$. The $0$-label informs us that all background quantities are evaluated at the same (conformal) time $\tau_0$.  To evaluate the amplitude, we need to choose a pivot scale, which constitutes a reference scale with which any other scale may be compared to. One alternative consists in choosing
\be
k_s = \frac{a_0 H_0}{c_0},
\ee
which is referred to as the sound horizon crossing condition. With this choice, $k = k_s$ labels the mode that had a wavelength that coincided with the sound horizon $c_0 / H_0$ at conformal time $\tau_0$. The amplitude of the power spectrum for that precise mode is then given by:
\be
\P_\R (k_s)  =  \frac{H_0^2 }{8  \pi^2 \epsilon_0 c_0} \bigg[ 1 +(2 \epsilon_0 + \eta_0 + s_0) \C  - 2 (\epsilon_0 + s_0)  \bigg] , \label{P-R-sound-horizon}
\ee
On the other hand, the spectral index evaluated at the pivot scale is
\be
n_{\R} - 1 \equiv \frac{d \ln \mathcal P_{\R}}{d \ln k} \bigg|_{k_s} = - ( 2 \epsilon_0 + \eta_0 + s_0 ) + \O (\epsilon_0^2) , \label{n-R-1}
\ee
where $\O (\epsilon_0^2) $ denotes contributions that are of second order in the slow-roll parameters. A similar derivation may be carried out to deduce the form of the tensor power spectrum. One finds, again, to first order in the slow-roll parameters, that this quantity is given by:
\be
\P_h (k)  =  \frac{2 H_h^2 }{\pi^2} \bigg[ 1 - 2 \epsilon_h  \ln \left( \frac{k }{a_h H_h} \right)  + 2 \epsilon_h ( \C - 1)    \bigg] . \label{P-h-horizon}
\ee
This time, we have used the label $h$ to denote that background quantities are evaluated at a time $\tau_h$ not necessarily equal to $\tau_0$. To correctly compare quantities, we need to evaluate the tensor spectrum $\P_h (k) $ at the same pivot scale $k_s$ as we did with the scalar spectrum. To do this, it is convenient to adjust $\tau_h$ in such a way that $k_s = a_h H_h$, or equivalently:
\be
\frac{H_h}{H_0} = \frac{a_0 }{a_h c_0} . \label{H-H-a-a}
\ee
This relation reminds us of the fact that if $c_0 \neq 1$, then scalar and tensor modes crossed the horizon at different times $\tau_0$ and $\tau_h$. We shall examine this statement in more details in a moment. To continue, the amplitude of the tensor power spectrum for the tensor mode of comoving wavelength $k = k_s$ is
\be
\P_h (k_s)  =  \frac{2 H_h^2 }{\pi^2} \bigg[ 1 + 2 \epsilon_h ( \C - 1)    \bigg] ,
\ee
whereas the spectral index of tensor perturbations is given by
\be
n_t \equiv \frac{d \ln \mathcal P_h }{d \ln k} \bigg|_{k_s} = - 2 \epsilon_h + \O(\epsilon_h^2). \label{n_t-def}
\ee
We may now compute the tensor to scalar ratio evaluated at $k = k_s$, which is given by $r = \P_h (k_s) / \P_\R (k_s)$, and reads:
\bea
r &=& 16 \epsilon_0 c_0 \left( \frac{H_h}{H_0} \right)^2 \bigg[ 1 + 2 \epsilon_h ( \C - 1)  \nn \\
&& - (2 \epsilon_0 + \eta_0 + s_0) \C  + 2 (\epsilon_0 + s_0)  \bigg]  . \label{r-first-exp}
\eea
To obtain a useful expression for the tensor to scalar ratio we need to make sense of quantities evaluated at the different conformal times $\tau_0$ and $\tau_h$. To proceed, let us go back to eq.~(\ref{H-H-a-a}) and count the number of $e$-folds $\Delta N = N_h - N_0$ between the two horizon exits (recall that $e$-folds are defined as $N \equiv \ln a$). One finds that:
\be
\Delta N = \ln \left( \frac{H_0}{H_h} \right) - \ln c_0 . \label{Delta-N-1}
\ee
On the other hand, because of eq.~(\ref{def-epsilon}), we see that
\be
\ln \left( \frac{H_0}{H_h} \right) = - \int^{N_0}_{N_h} \!\!\! \epsilon(N) dN = \frac{\epsilon_0}{\eta_0} \left[ e^{\eta_0 \Delta N}  -  1  \right] , \label{Delta-N-2}
\ee
where we have used the fact that $\epsilon (N) = \epsilon_0 e^{\eta_0 (N - N_0)}$, as long as we treat $\eta_0$ as a constant (which is justified since the running of $\eta$ only contributes higher order effects in slow-roll). Putting together eqs.~(\ref{Delta-N-1}) and~(\ref{Delta-N-2}), we find:
\be
\Delta N + \ln c_0 = \frac{\epsilon_0}{\eta_0} \left[  e^{\eta_0 \Delta N} - 1\right] . \label{Delta-N-3}
\ee
We immediately see that $\Delta N \sim - \ln c_0$. However, given that $- \ln c_0$ may attain large values in the range $c_0 \ll 1$, we are forced to admit the possibility that $- \epsilon_0 \ln c_0$ and $- \eta_0 \ln c_0$ could both reach values of order $1$ (without implying a violation of the slow-roll conditions). This implies that we cannot expand the exponential of eq.~(\ref{Delta-N-3}) in powers of $\eta_0 \Delta N$ to derive a simple expression for $\Delta N$. Despite of this, we can analytically solve eq.~(\ref{Delta-N-3}) to obtain:
\be
\Delta N = - \ln c_0 - \frac{1}{\eta_0} \left[ \epsilon_0 + W \! \left(- \epsilon_0 e^{- \epsilon_0 - \eta_0 \ln c_0} \right) \right] , \label{Delta-N-W-1}
\ee
where $W(x)$ is the Lambert-$W$ function, defined as the solution of the equation $x = W(x) e^{W(x)}$. Putting all of these results together back in eq.~(\ref{r-first-exp}), we finally obtain
\bea
r &=& 16 \epsilon_0 c_0 e^{-2 ( \Delta N + \ln c_0)} \bigg[ 1 + 2 \epsilon_0 e^{\eta_0 \Delta N} ( \C - 1)  \nn \\
&& - (2 \epsilon_0 + \eta_0 + s_0) \C  + 2 (\epsilon_0 + s_0)  \bigg]  , \label{r-second-exp}
\eea
where we have used the additional relation between $\epsilon_h$ and $\epsilon_0$:
\be
\epsilon_h = \epsilon_0 e^{\eta_0 \Delta N}. \label{epsilon-h-epsilon-0}
\ee
Equation (\ref{r-second-exp}) gives us $r$ as a function of  $\epsilon_0$, $\eta_0$, $c_0$ and $s_0$. However, we may reduce the number of parameters entering this expression by using eq.~(\ref{n-R-1}) and introducing the observed value of the spectral index $1 - n_\R = 0.04$~\cite{Ade:2013uln, Ade:2015lrj}. This means that $\Delta N$ is a function of $c_0$, $\epsilon_0$ and $s_0$ given by
\bea
\Delta N &=& -  \ln c_0 - \frac{1}{1 - n_\R - 2 \epsilon_0 - s_0} \bigg[ \epsilon_0 \qquad \qquad  \nn \\
&& + W \! \left(- \epsilon_0 e^{- \epsilon_0 - (1 - n_\R - 2 \epsilon_0 - s_0) \ln c_0} \right) \bigg] , \label{Delta-N-W-2}
\eea
with the understanding that the value of $n_\R$ is fixed by observations.

\subsection{Adding a measurement of the tensor spectral index $n_t$}

Let us recall that $n_t$  gives us the value of $\epsilon_{h}$, which differs from $\epsilon_0$ in the event that the sound speed $c_s$ is much smaller than $1$. In fact, putting together eqs.~(\ref{epsilon-h-epsilon-0}) and~(\ref{Delta-N-3}), we see that a measurement of $n_t$ would reduce the number of parameters that $\Delta N$ depends on, giving us:
\be
 \Delta N  = -  \frac{\ln c_0}{1- ( \epsilon_0 + n_t/2 )/\ln \left( - \frac{2 \epsilon_0}{n_t} \right)}  . \label{DeltaN-nT}
\ee
This relation allows us to reduce the dependence of $r$ down to $\epsilon_0$ and $c_0$, returning
\be
r = 16 \epsilon_0 c_0 \exp \bigg[ \frac{-2 \ln c_0}{1 - \ln \left( - \frac{2 \epsilon_0}{n_t} \right) \big/ (  \epsilon_0 + n_t/2) } \bigg] \left( 1 + \cdots \right) ,\label{r-nT}
\ee
where the elipses $\cdots$ stand for the same slow roll corrections of eq.~(\ref{r-second-exp}), that may also be expressed in terms of $\epsilon_0$ and $c_0$ (and the observed values of $n_\R$ and $n_t$). Eq.~(\ref{r-nT}) is one of our main results. It gives the dependence of $r$ in terms of the parameters $\epsilon_0$ and $c_0$ provided that $n_\R$ and $n_t$ are known.

\section{Constraints on the sound speed} \label{sec:constraints}

\begin{figure}[t!]
\includegraphics[scale=0.4]{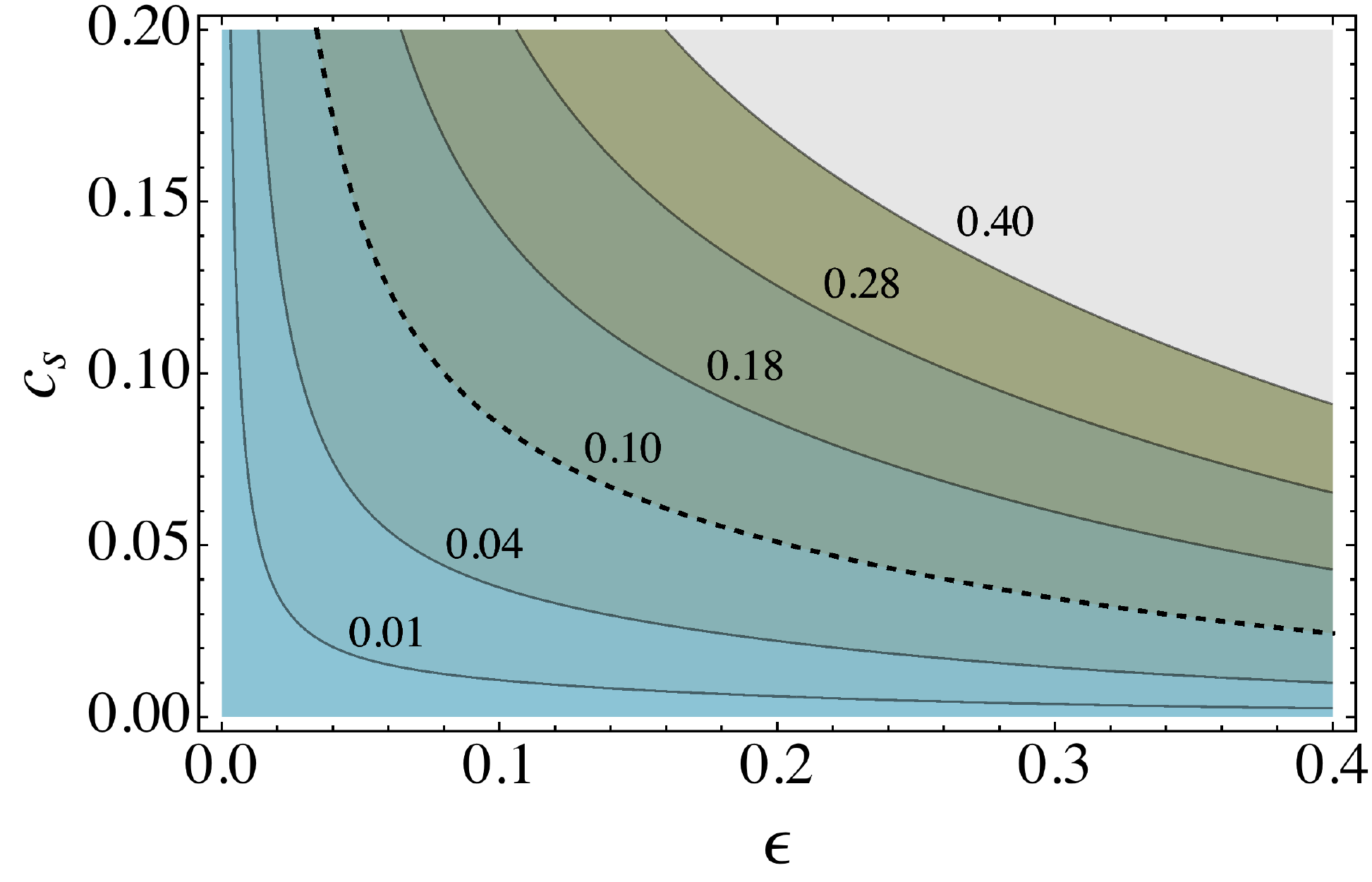}
\caption{The figure shows the contour plot for $r$ in the $\epsilon$-$c_s$ plane, obtained from eq.~(\ref{r-second-exp}) assuming that $s_0 = 0$. The dashed line shows the case for $r = 0.1$.}
\label{fig:ec-no-s}
\end{figure}

We now examine the constraints on the possible values of the sound speed $c_s$ implied by eq.~(\ref{r-second-exp}). We first consider the simple case in which the sound speed $c_s$ remains constant, that is $s = 0$, and then move on to consider how these constraints are affected by additional considerations (such as the constraint on the running of the spectral index). FIG.~\ref{fig:ec-no-s} shows the contour plots for $r$ in the $\epsilon$-$c_s$ plane obtained from eq.~(\ref{r-second-exp}) after  replacing $s_0 = 0$ in eq.~(\ref{Delta-N-W-2}). The case $r=0.1$ has been highlighted by the dashed contour.
We see that this plot differs substantially from the one shown in FIG.~\ref{fig:ec-basic-log} based on eq.~(\ref{basic-r-corr}), and does not imply a lower bound on $c_s$. This is due to the non vanishing value of $\eta_0$ affecting the running of $\epsilon$ from the scalar horizon exit time $\tau_0$ to the tensor horizon exit time $\tau_h$. As emphasized in~\cite{Baumann:2014cja}, to obtain a bound on $c_s$ taking into account a non-vanishing value of $\eta_0$ requires one to consider the additional constraint on the running of the scalar spectral index $\alpha$. We examine this point in Section~\ref{sec:running-alpha}.

\subsection{The consistency relation} \label{sec:consistency-relation}

\begin{figure}[t!]
\includegraphics[scale=0.4]{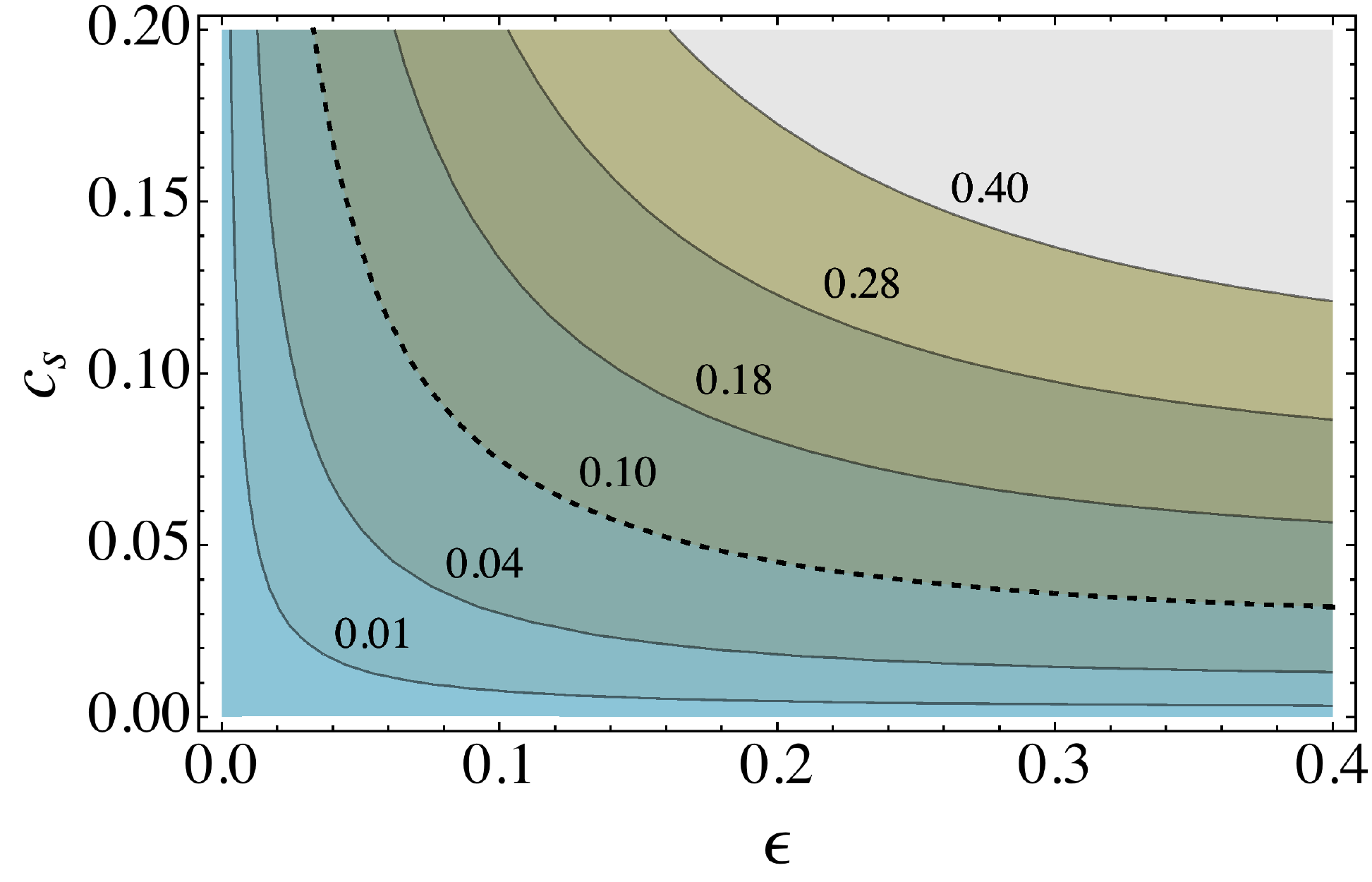}
\caption{The figure shows the contour plot for $r$ in the $\epsilon$-$c_s$ plane, obtained from eq.~(\ref{r-nT}) satisfying the consistency relation $r = - 8 n_t$. The dashed line shows the case for $r = 0.1$.}
\label{fig:ec-r-nt}
\end{figure}

In canonical models of inflation ($c_s = 1$) the tensor to scalar ratio $r$ and the spectral index of tensor modes $n_t$ reduce to 
\be
r = 16 \epsilon_0, \qquad n_t = - 2 \epsilon_0 .
\ee
These results lead to the well known consistency relation:
\be
r = - 8 n_t. \label{consist-1}
\ee
As we have seen, in the case of non-canonical models of inflation, the tensor to scalar ratio $r$ may be written in terms of $\epsilon_0$, $c_0$ and $n_t$ as in eq.~(\ref{r-nT}). On the other hand, the tensor spectral index is determined by the value of $\epsilon$ at the time tensor horizon crossing (that is $\epsilon_h$) as expressed in eq.~(\ref{n_t-def}). This implies that the consistency relation (\ref{consist-1}) may be satisfied even for $c_0 \neq 1$ as long as the following relation is satisfied: 
\be
- \frac{2 \epsilon_0}{ n_t} c_0 \exp \bigg[ \frac{2 \ln c_0}{1 - \ln \left( - \frac{2 \epsilon_0}{n_t} \right) \big/ (  \epsilon_0 + n_t/2) } \bigg]  = 1 .  \label{consist-2}
\ee
In other words, a measurement of the tensor spectral index does not eliminate the degeneracy between $\epsilon$ and $c_s$ as usually thought. The origin of this degeneracy is the running of $\epsilon$ between the two horizon crossing times. Notice that if there is no running of $\epsilon$ between the two horizon exits, one has $\epsilon_h = \epsilon_0$ (that is $\eta_0 = 0$) from where one sees that $n_t = - 2 \epsilon_0$. Plugging this back into (\ref{consist-2}) one obtains $c_0 =1$, implying no distinction between the two horizon crossing times. FIG.~\ref{fig:ec-r-nt} shows the allowed values of $c_0$ and $\epsilon_0$ that satisfies the consistency relation~(\ref{consist-1}) for various values of $r$, which marginally differs from the contour plots of FIG~\ref{fig:ec-no-s}.
On the other hand, it should be noticed that in order to satisfy the consistency relation~(\ref{consist-1}), one requires that the speed of sound has a running respecting $s_0 = (1 - n_\R) - 2 \epsilon_0 - \eta_0$, coming from eq.~(\ref{n-R-1}). This implies:
\be
s_0 = (1 - n_\R) - 2 \epsilon_0 + \frac{1}{\Delta N} \ln \left( -\frac{2 \epsilon_0}{n_t} \right)
\ee
where $\Delta N$ is given by (\ref{DeltaN-nT}). In Section~\ref{sec:running-alpha} we further consider the effects of the running of the scalar spectral index $\alpha$ on this analysis.

\subsection{Running sound speed} 
 
\begin{figure}[t!]
\includegraphics[scale=0.4]{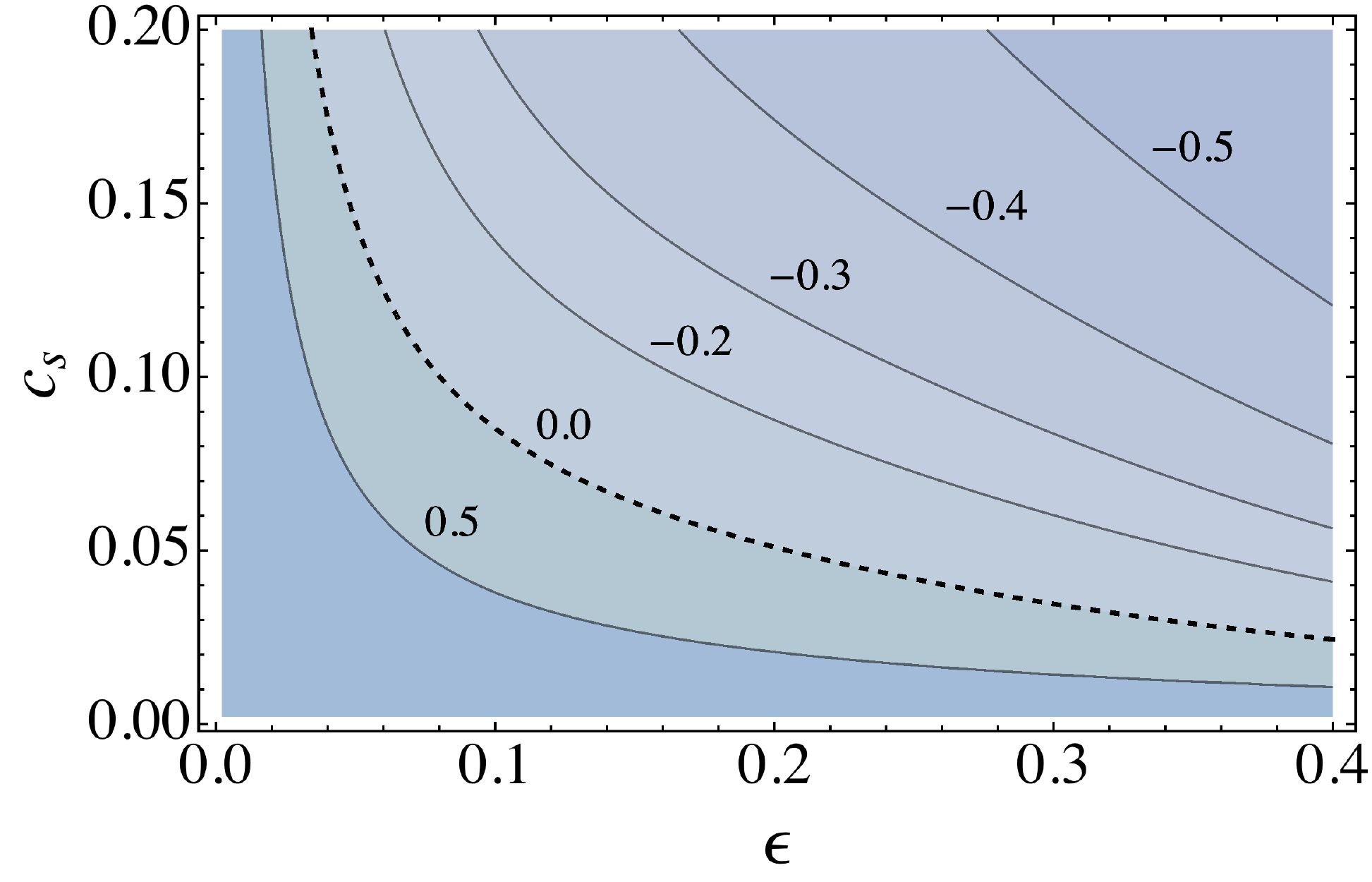}
\caption{The figure shows the contour plot for $s_0$ in the $\epsilon$-$c_s$ plane, obtained from eq.~(\ref{r-second-exp}) for the particular case $r = 0.1$. The dashed line shows the case for $s_0 = 0$.}
\label{fig:ec-s}
\end{figure}

\begin{figure}[t!]
\includegraphics[scale=0.4]{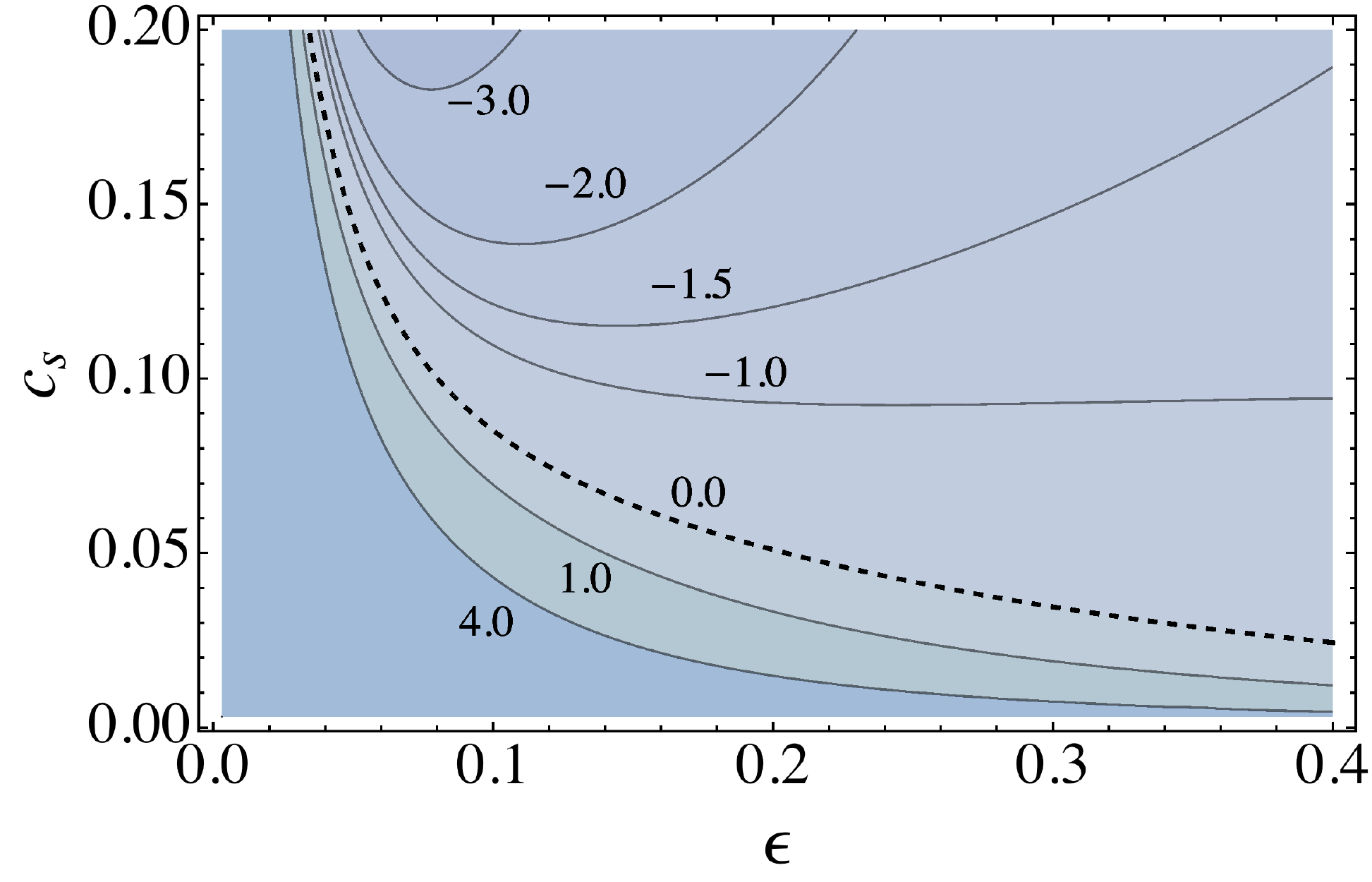}
\caption{The figure shows the contour plot for $\lambda$ in the $\epsilon$-$c_s$ plane, obtained by putting together eqs.~(\ref{r-second-exp}) and~(\ref{s-lambda}), for the particular case $r = 0.1$. The dashed line corresponds to the case $\lambda = 0$.}
\label{fig:ec-lambda}
\end{figure}

Let us now consider the case in which the sound speed $c_s$ is allowed to evolve, parametrized by a non-vanishing value of $s$. FIG.~\ref{fig:ec-s} shows the contour plot for $s$ in the $\epsilon$-$c_s$ plane, in the particular case $r = 0.1$. The dashed curve corresponds to $s=0$.
We see that the presence of a fixed non-vanishing value of $s$ does not introduce a drastic change on the relation between $\epsilon$ and $c_s$. A negative running of the sound speed ($s<0$) tends to increase the value of $c_s$ for a fixed value of $\epsilon$. Next, we may consider the realistic possibility in which $s$ depends on $\epsilon$. For instance, the authors of ref.~\cite{Achucarro:2012yr} studied a multi-field model where the inflaton trajectory consisted of an almost constant turn that, thanks to the interaction between the inflaton and heavy fields ortogonal to the trajectory, the sound speed of adiabatic perturbations was characterized by a running of the form $s = - \epsilon / 4$. Motivated by this example, we choose to model the dependence of $s$ on $\epsilon$ by the following simple parametrization:
\be
s(\epsilon) = \lambda \epsilon, \label{s-lambda} 
\ee
where $\lambda$ is a constant. FIG.~\ref{fig:ec-lambda} shows the contour plot for $\lambda$ in the $\epsilon$-$c_s$ plane in the particular case $r = 0.1$. The dashed curve corresponds to $\lambda=0$.
We see that the dependence of $c_s$ on $\epsilon$ is drastically affected by $\lambda$. In this particular example, where $r=0.1$, we find that for values $\lambda \lesssim -1.2$ one recovers a lower bound on $c_s$ given by $c_s \ge 0.9$. Thus, as a general rule, we find that a negative running implies stronger lower bounds on $c_s$ for a fixed value of $\epsilon$. Moreover, if the running is proportional to $\epsilon$, one finds configurations with lower bounds on $c_s$.

\subsection{Including the running of the spectral index $\alpha$}  \label{sec:running-alpha}

\begin{figure}[t!]
\includegraphics[scale=0.4]{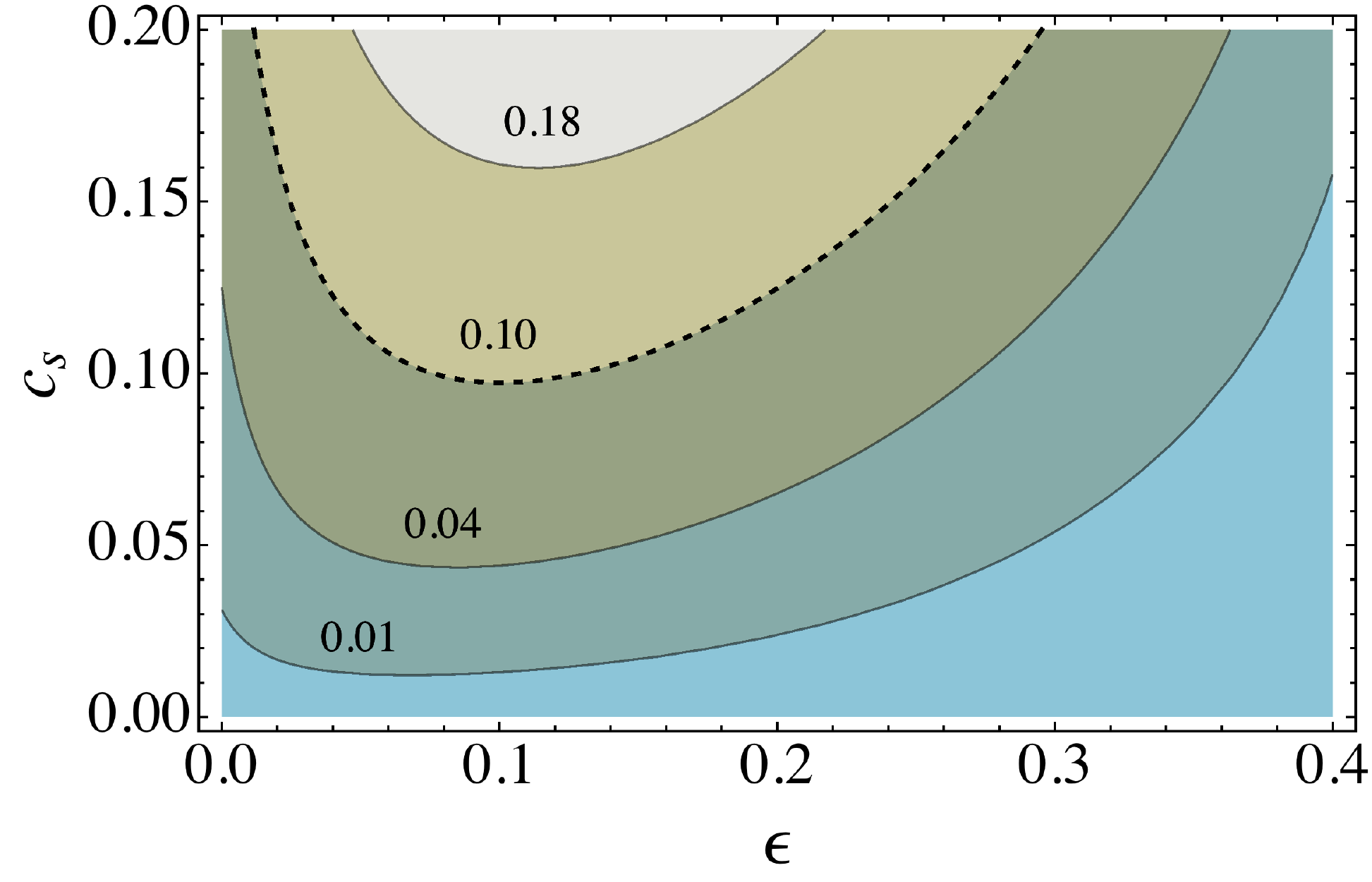}
\caption{The figure shows the contour plot for $r$ in the $\epsilon$-$c_s$ plane, obtained from eq.~(\ref{r-second-exp}) taking into account the constraints on the running $\alpha$. The dashed line shows the case for $r = 0.1$.}
\label{fig:ec-alpha}
\end{figure}

A crucial aspect of the analysis performed in ref.~\cite{Baumann:2014cja} was to take into account current constraints on the running $\alpha$ of the spectral index $n_\R - 1$ of scalar perturbations. It is customary to define the running $\alpha$ through the following parametrization of the scalar power spectrum
\be
\P_{\R}(k) = \P_0 \left( \frac{k}{k_*} \right)^{n_\R - 1 + \frac{1}{2} \alpha \ln (k / k_*) + \cdots} ,
\ee
where $k_*$ is a pivot scale. Then, $\alpha$ is found to have the following dependence on other slow roll parameters
\be
\alpha = - 2 \epsilon  \eta - \eta  \delta_\eta - s  \delta_s   , \label{alpha}
\ee
where $\delta_\eta$ and $\delta_s$ are slow roll parameters required to satisfy $|\delta| \ll 1$, and defined as:
\be
\delta_\eta = \frac{\dot \eta}{H \eta} , \qquad \delta_{s} = \frac{\dot s}{H s} .
\ee
Current observations impose the constraint~\cite{Ade:2013uln}:
\be
|\alpha| \leq 2 \times 10^{-2}.  \label{alpha-constr}
\ee
(See also~\cite{Ade:2015lrj} for the latest update from Planck, which slightly modify this constraint). Given that $\alpha$ depends on many slow-roll parameters (\ref{alpha}), in principle, it is possible to satisfy (\ref{alpha-constr}) in several ways by conveniently adjusting $\delta_\eta$ and $\delta_s$. However, as observed in~\cite{Baumann:2014cja}, in order to respect the hierarchical structure involved in the slow-roll expansion, the main contribution to~(\ref{alpha}) must be given by the first term:
\be
\alpha \simeq - 2 \epsilon \, \eta  .  \label{alpha-2}
\ee
Combining this expression with (\ref{alpha-constr}), one obtains:
\be
\epsilon_0 \, |\eta_0| < 10^{-2} . \label{eta-constr} 
\ee
Then, by setting $ \eta_0 = 10^{-2} / \epsilon_0$ in eqs.~(\ref{r-second-exp}) and~(\ref{Delta-N-W-1}) one obtains new restrictions on the possible values of $c_s$ and $\epsilon$. The resulting contour plot for $r$ in the $\epsilon$-$c_s$ plane is shown in FIG.~
\ref{fig:ec-alpha}, where it is possible to see that for the reference value $r=0.1$, the speed of sound has a lower bound $c_s > 0.1$~\footnote{Notice that now the contour lines intersect with $\epsilon_0 = 0$. This comes from using the relation $ \eta_0 = 10^{-2} / \epsilon_0$ which invalidates the use of eq.~(\ref{r-second-exp}) at values of $\epsilon$ of order $10^{-2}$.}.
On the other hand, the combined data sets from BICEP2 and Planck tend to favor a negative running $\alpha$, which in turn, imposes stronger constraints on $c_s$. For instance, following the analysis of ref.~\cite{Kinney:2014jya}, we obtain the bounds $c_s > 0.25$ at 68\% confidence level, and $c_s > 0.2$ at 95\% confidence level, corroborating the bound of ref.~\cite{Baumann:2014cja}.

To finish this section, let us examine how eq.~(\ref{alpha-2}) affects our determination of $c_s$ in the case where $n_t$ is already known. Putting together eqs.~(\ref{epsilon-h-epsilon-0}),~(\ref{DeltaN-nT}) and~(\ref{alpha-2}), we see that
\be
\alpha =   - \frac{2 \epsilon_0}{\ln c_0} \left[ \ln \left( -\frac{2 \epsilon_0}{n_t} \right) - ( \epsilon_0 + n_t/2 ) \right] .
\ee
This relation allows us to break the degeneracy between $c_0$ and $\epsilon_0$ provided that $\alpha$ is known (and assuming that $\delta_\eta$ and $\delta_s$ are negligible). 

\section{Parametrizing additional UV-physics} \label{sec:uv-par}

As already emphasized in the introduction, once we accept the possibility of having $c_s \neq 1$, we should admit a variety of additional operators proportional to $1 - c_s^2$ that have their origin on the same UV physics responsible for a non-trivial propagation of adiabatic perturbations~\cite{ArkaniHamed:2003uz, Cheung:2007st, Baumann:2011su, Achucarro:2012yr, Gwyn:2012mw}.  The aim of this section is to include the effects of such UV degrees of freedom on the bounds of $c_s$. One way of proceeding is to parametrize the effects of UV physics by including new operators in the quadratic action of adiabatic curvature fluctuations $\R$, in the following manner
\be
S_{\rm UV} \!\! = \!\! \int \!\! d^4x  a^3 \frac{\epsilon \left( 1 - c_s^2 \right)}{M^2} \left[ \frac{\beta_1}{a^2 c_s^2}  \dot {\mathcal R} \nabla^2 \dot {\mathcal R}  +  \frac{c_s^2 \beta_2}{a^4}  {\mathcal R} \nabla^4  {\mathcal R}  \right] ,  \label{S-UV}
\ee
where $M$ is a mass scale characterizing the new degrees of freedom, but not necessarily the cutoff scale at which they become operative~\cite{Achucarro:2012yr}, and $\beta_1$ and $\beta_2$ are dimensionless coefficients parametrizing the UV physics. We have inserted the global factor $1 - c_s^2$, to incorporate the fact that one should recover single field canonical inflation in the limit $c_s \to 1$. It is then straightforward to deduce that the interaction picture hamiltonian taking into account these new operators is given by
\bea
H_I \!\! &=& \!\! - \frac{1 - c_s^2 }{2 a^2 M^2} \! \int_x  \! \left[  \beta_1 u_I' \nabla^2 u_I' + c_s^4 \beta_2   u  \nabla^4 u  \right] , \label{uv-hamiltonian}
\eea
where $\int_x \equiv \int d^3 x$, $u = z \R$ and $z = a \sqrt{\epsilon} / c_s$. Using the results of Appendix~\ref{app-pert} it is then possible to deduce that the scalar power spectrum receive new contributions given by 
\be
\frac{\Delta \P_{\rm UV} }{\P_0}= \frac{\left( 1 - c_s^2 \right) H_0^2}{ 4 c_0^2 M^2} \left( \beta_1 + 5 \beta_2 \right) .
\ee
Thus, we see that the leading effect of these new terms is to modify the amplitude of the scalar power spectrum, which now is given by
\bea
\mathcal P_\R (k)  =  \frac{H_0^2 }{8  \pi^2 \epsilon_0 c_0} \bigg[ 1 - (2 \epsilon_0 + \eta_0 + s_0) \ln \left( \frac{ c_0 k}{a_0 H_0} \right)  \nn \\
 +(2 \epsilon_0 + \eta_0 + s_0) \C  - 2 (\epsilon_0 + s_0) +  \frac{1-  c_0^{2} }{4 c_0^{2}} \beta   \bigg] ,\label{P-R-beta}
\eea
where $\beta \equiv (\beta_1 + 5 \beta_2) H_0^2 / M^2$. On the other hand, tensor modes are not affected by the new terms in eq.~(\ref{S-UV}), and we are allowed to use eq.~(\ref{P-h-horizon}) to characterize them. We may now proceed to compute the tensor to scalar ratio using the same procedure of Section~\ref{sec:tensor-scalar}, to obtain
\bea
r &=& 16 \epsilon_0 c_0 e^{-2 ( \Delta N + \ln c_0)} \bigg[ 1 + 2 \epsilon_0 e^{\eta_0 \Delta N} ( \C - 1)  \nn \\
&& - ( 1 - n_\R ) \C  + 2 (\epsilon_0 + s_0) -  \frac{1-  c_0^{2} }{4 c_0^{2}} \beta  \bigg]  , \label{r-second-exp-beta}
\eea
where $\Delta N$ is given by (\ref{Delta-N-W-2}). It is worth mentioning that models of inflation with a single heavy field of mass parameter $M$, interacting with curvature perturbations corresponds to the particular case $\beta_1 = 1$ and $\beta_2 = 0$, in the limit $H_0^2 \ll M^2$~\cite{Achucarro:2012sm, Cespedes:2013rda}. Because the term proportional to $\beta$ contains the factor $c_0^{-2} - 1$, eq.~(\ref{r-second-exp-beta}) tells us that even a very small value of the parameter $\beta$ can have a large impact on the dependence of $r$ on the sound speed $c_s$. This term quickly dominates the square bracket in eq.~(\ref{r-second-exp-beta}) at small values of $c_0$, implying a breakdown of our expansion. Despite of this, we may draw the following simple conclusion: A positive value of $\beta$ implies much stronger lower bounds on $c_s$ for a fixed value of $\epsilon$.

\section{Conclusions} \label{sec:conclusions}

A measurement of a large value of the tensor to scalar ratio $r$ would constitute a dramatic breakthrough in our understanding of the very early universe. In this article, we have analyzed the implications of a large value of $r$ on the evolution slow-roll quantities parametrizing deviations from canonical single field inflation. Our work has been greatly motivated by ref.~\cite{Baumann:2014cja}, where it was emphasized that a small value of the sound speed could have a sizable impact on the dependence of $r$ on the running of background inflationary parameters. Our results ratify this assertion. More importantly, we have seen that the same effects leading to the constraints on the speed of sound, preserve the degeneracy between $\epsilon$ and $c_s$ even in the case where the consistency relation is found to be satisfied. As we have seen in Section~\ref{sec:consistency-relation}, a measurement of the spectral index of tensor modes $n_t$ does not break the degeneracy between $\epsilon$ and $c_s$, and a confirmation of the consistency relation does not necessarily rule out non-canonical models of inflation. However, a determination of the running of the scalar spectral index $\alpha$ would improve substantially our knowledge about non-canonical models parametrized by $c_s$. Our results emphasize the importance CMB-polarization experiments~\cite{Baumann:2008aq} in order to constrain nontrivial deviations from canonical single field inflation (parametrized by the speed of sound and the UV-physics parameters $\beta_1$ and $\beta_2$). More precisely, if future CMB experiments~\cite{EssingerHileman:2010hh, Kermish:2012eh, Abazajian:2013vfg, CLASS} detect a signal larger than $r=0.01$ we would count with better constraints on the value of $c_s$ (and even on the UV-physics parameter $\beta$) than those obtained from non-Gaussianity observations.

\begin{acknowledgements}

We are grateful to Sander Mooij for useful comments and discussions. 
This work was supported by the Fondecyt project 1130777 (GAP \& AS), an Anillo project ACT1122 (GAP \& AS), and a Conicyt Fellowship CONICYT-PCHA/MagisterNacional/2013-221320624 (AS). \\

\end{acknowledgements}
\vspace{-20pt}

\begin{appendix}

\section{Perturbation theory} \label{app-pert}

In this appendix we review the in-in formalism of perturbation theory applied to compute $2$-point correlation functions in inflationary backgrounds~\cite{Maldacena:2002vr, Weinberg:2005vy}. 

\subsection{Slowly-rolling background} 

Let us start by considering the quadratic action parametrizing the evolution of adiabatic perturbations $\R (\x , t)$ with a nontrivial sound speed $c_s$. This action is given by
\be
S =  \int d^3x \, dt  a^3 \epsilon \left[ \frac{1}{c_s^2} \dot {\mathcal{R}}^2 - \frac{1}{a^2}(\nabla \mathcal{R})^2  \right] ,
\ee
where $a$ is the scale factor and where $\epsilon$ is the usual slow-roll parameter given by:
\be
\epsilon \equiv - \frac{\dot H}{H^2} , \qquad H \equiv \frac{\dot a}{a}.
\ee
Notice that we are working in units such that $m_{\rm Pl}^2=1$.  We may parametrize the time evolution of $\epsilon$ and $c_s$ through additional slow-roll parameters $\eta$ and $s$ defined as:
\be
\eta \equiv \frac{\dot \epsilon}{H \epsilon} , \qquad s = \frac{\dot c_s}{H c_s} .
\ee
Since we are interested computing the power spectrum of scalar modes to first order in the slow-roll parameters, there is no need to further parametrize the evolution of $\eta$ and $s$, which would lead to second order effects. Then, taking $\eta$ and $s$ to be constant, we obtain
\bea
\epsilon &=& \epsilon_0  e^{ \eta_0 \ln (a / a_0) }  ,  \\ 
c_s &=& c_0  e^{ s_0 \ln (a / a_0) }  ,
\eea
where $\epsilon_0$ and $c_0$ are the values of $\epsilon$ and $c_s$ at a reference time $t_0$, when the scale factor is given by $a(t_0) = a_0$. To simplify the computation we may expand the exponentials as
\bea
\epsilon &=& \epsilon_0 \left[ 1 + \eta_0 \ln (a / a_0)+  \cdots \right] ,  \label{epsilon-a} \\ 
c_s &=& c_0 \left[ 1 + s_0 \ln (a / a_0)  + \cdots \right] .
\eea
Integrating (\ref{epsilon-a}) we obtain $H$ as
\be
H = H_0 \left[ 1 - \epsilon_0 \ln (a / a_0)+ \frac{\epsilon_0 (\epsilon_0 - \eta_0)}{2} \ln^2 (a / a_0) + \cdots \right] ,
\ee
where $H_0$ is the value of the Hubble parameter evaluated at  time $t_0$.  It is convenient to work with conformal time $\tau$, which comes defined through the change of variables $d t = a d \tau$. Then, integrating one more time we obtain an expression for the scale factor $a$ as a function of conformal time $\tau$
\be
a = a_0(\tau) \left[ 1 + A_1 + A_2  + \cdots \right] , \quad a_0(\tau) = - \frac{a_0}{H_0 \tau} , 
\ee
where 
\bea
A_1 (\tau)  &=& \epsilon_0 \left[ 1 + \ln (a_0(\tau)/a_0) \right] , \\
A_2 (\tau)  &=& \frac{1}{2}  \epsilon_0 \bigg[ 2 (2 \epsilon_0  + \eta_0) + 2 (2 \epsilon_0 + \eta_0) \ln (a_0(\tau)/a_0)  \nn \\
&&  + (\epsilon_0 + \eta_0 ) \ln^2 (a_0(\tau)/a_0)  \bigg] .
\eea
Notice that $a_0(\tau) = a_0$ at a conformal time given by:
\be
\tau_0 = - \frac{a_0}{H_0} \left[ 1 + \epsilon_0 + \epsilon_0 (\epsilon_0 + \eta_0)  + \cdots  \right]  .
\ee
In what follows we simplify our computations by setting $a_0 =1$. We will later restore the value $a_0$ whenever it becomes necessary. To deal with the dynamics of perturbations it is convenient to define a canonical field $u$ through the following rescaling of $\R$:
\be
u = z \mathcal{R}, \qquad z = \sqrt{2 \epsilon} \frac{a}{c_s} . 
\ee
Then, the action for the $u$-field is found to be given by
\be
S =  \frac{1}{2} \int d^3x \, d \tau  \left[ (u')^2 - c_s^2 (\nabla u)^2 + \frac{z''}{z} u^2 \right] , \label{S-u}
\ee
where the prime $'$ denotes derivatives with respect to conformal time $\tau$. Expanding the coefficient $z''/z$ up to first order in slow-roll, we obtain:
\bea
 \frac{z''}{z} &=& \frac{2}{\tau^2} + \frac{3 (2 \epsilon_0 + \eta_0 - 2 s_0)}{2 \tau^2}  + \cdots .
\eea
This expression implies a natural splitting of the theory between a zeroth order part, and a first order part in terms of the slow-roll parameters $S = S_0 + S_1$, where:
\bea
S_0 \!\! &=& \!\!  \frac{1}{2} \int \! d^3x \, d \tau  \left[ (u')^2 - c_0^2 (\nabla u)^2 + \frac{2}{\tau^2} u^2 \right] , \label{S-0} \\
S_1 \!\! &=& \!\! \frac{1}{2}\int \! d^3x \, d \tau  \left[  - c_0^2  \theta ( \tau )  (\nabla u)^2 + \frac{1}{\tau^2}   \delta_0 \,  u^2 \right], \label{S-1}
\eea
where we have defined
\bea
\delta_0  &=& \frac{3}{2} (2 \epsilon_0 + \eta_0 - 2 s_0), \\
\theta_0 (\tau) &=&  2  s_0 \ln a_0(\tau) .
\eea
The splitting $S = S_0 + S_1$ will allow us to compute the power spectrum with the help of perturbation theory to first order in slow-roll.

\subsection{Perturbation theory}

The interaction piece of eq.~(\ref{S-1}) defines the Hamiltonian of the interaction picture as
\be
H_{\rm I} (\tau) = \frac{1}{2} \int d^3x  \left[  c_0^2  \theta ( \tau )  (\nabla u_I)^2 - \frac{\delta_0 }{\tau^2} u_I^2 \right] ,
\ee
where $u_I$ is the interaction picture field defined as
\be
u_I = \frac{1}{(2 \pi)^3} \int d^3 k \left[ a_\k u_k(\tau) e^{i \k \cdot \x}  + a_\k^{\dag} u_k^{*}(\tau) e^{-i \k \cdot \x}   \right] ,
\ee
where the pair $a_\k^\dag$ and $a_\k$ are the usual creation and annihilation operators satisfying the commutation relation
\be
\big[ a_\k , a_{\k'}^\dag \big] = (2 \pi)^3 \delta^{(3)} (\k - \k'),
\ee
and $u_k(\tau)$ represents the normalized solution to the zeroth order equation of motion deduced from (\ref{S-0}), given by:
\be
u_k(\tau) = \frac{1}{\sqrt{2 c_0 k}} \left( 1 - \frac{i}{c_0 k \tau} \right) e^{- i c_0 k \tau} .
\ee
Standard perturbation theory tells us that the complete solution $u(x,\tau)$ may be written in terms of the interaction picture field $u_I(x,\tau)$ and the propagator $U(\tau)$ as
\be
u(\x,\tau) = U^{\dag}(\tau)  u_I(\x,\tau) U(\tau) ,
\ee
with 
\be
U(\tau) = \mathcal T \exp \left\{ - i \int^{\tau}_{-\infty_+} d\tau' H_I (\tau') \right\} ,
\ee
where $ \mathcal T$ stands for the time ordering symbol, and $\infty_+ = (1 + i \epsilon) \infty$ is the usual prescription to isolate the in-vacuum. We may now compute the two-point correlation function for $u$, which is written in terms of the interaction picture quantities as:
\be
\langle u(x, \tau) u(y, \tau)  \rangle = \langle 0 | U^\dag (\tau)  u_I(x,\tau) u_I(y,\tau) U(\tau)   | 0 \rangle  .  
\ee
By expanding the previous result up to first order in $H_I$, we obtain:
\bea
\langle u(\x, \tau) u(\y, \tau)  \rangle =  \langle 0| u_I(x,\tau) u_I(y,\tau) | 0 \rangle \qquad \quad \nn \\
+ i \int^{\tau}_{-\infty} d\tau'   \langle 0|  \left[ H_I(\tau') ,u_I(\x,\tau) u_I(\y,\tau) \right] | 0 \rangle \label{correlator},
\eea
which is the two-point correlation function for the $u$-field up to first order in slow-roll.

\subsection{Scalar Power spectrum}

Let us now consider the computation of the power spectrum for adiabatic fluctuations $\mathcal P_{\mathcal R} ( k )$. Because $\R = u / z$, we find that $\mathcal P_{\mathcal R} ( k )$ is given in terms of the two-point correlation function for the $u$-field as
\be
\mathcal P_{\mathcal R} ( k , \tau) =  \frac{4 \pi k^3}{(2 \pi)^3 z^2} \int_y \! \langle u(\y,\tau) u(0,\tau)  \rangle  e^{- i \k \cdot  \y} , \label{power-spectrum-R-u}
\ee
where $\int_y$ stands for $\int \! d^3 y$. Expanding $z$ up to first order in slow-roll as $z = z_0 + z_1$, where 
\bea
& z_0 =  \frac{\sqrt{\epsilon_0} a_0(\tau)}{c_0} , \nn \\
& z_1 =  \frac{\sqrt{\epsilon_0} a_0(\tau)}{c_0} \left[ \epsilon_0 + (\epsilon_0 -  s_0 + \eta_0/2  ) \ln a_0(\tau) \right] , \quad 
\eea 
and using (\ref{correlator}) back into (\ref{power-spectrum-R-u}), we obtain:
\bea
\mathcal P_{\mathcal R} ( k , \tau) =  \frac{4 \pi k^3}{(2 \pi)^3 z_0^2} \int_y \! \langle 0 | u_I(\y,\tau) u_I(0,\tau)  | 0 \rangle  e^{- i \k \cdot  \y} \nn \\ 
- \frac{8 \pi k^3 z_1}{(2 \pi)^3 z_0^2}  \int_y \! \langle 0 | u_I(\y,\tau) u_I(0,\tau)  | 0 \rangle  e^{- i \k \cdot  \y} \nn \\ 
+ \frac{4 \pi i k^3}{(2 \pi)^3 z_0^2}    \int_y \! \int^{\tau}_{-\infty} \!\!\!\!\!  d\tau'   \langle 0|  \left[ H_I(\tau') ,u_I(\y ,\tau) u_I( 0 ,\tau) \right] | 0 \rangle . \nn \\ \label{complete-P}
\eea
The rest of the computation is straightforward: The first line of (\ref{complete-P}) gives the zeroth order power spectrum $\mathcal P_{\R}^{(0)} $, which is scale independent, whereas the two next lines combine to give the first order correction $\Delta \mathcal P_\R (k, \tau)$, which depends on the scale $k$ and on time $\tau$. In the long wavelength limit $|\tau| k \ll 1$ the final result has the form
\be
\mathcal P_{\mathcal R} (k)  = \mathcal P_{\R}^{(0)} + \Delta \mathcal P_\R (k) , 
\ee
where
\bea
\P_{\R}^{(0)} &=&  \frac{H_0^2 }{8  \pi^2 \epsilon_0 c_0}  , \\
\frac{ \Delta \mathcal P_\R (k) }{\P_{\R}^{(0)} }&=&- (2 \epsilon_0 + \eta_0 + s_0)  \ln ( c_0 k/ a_0 H_0) \nn \\
&& + (2 \epsilon_0 + \eta_0 + s_0) \C - 2 (  \epsilon_0 +  s_0 ) ,
\eea
where $\C = 2 - \log 2 - \gamma$ ($\gamma$ being the Euler-Mascheroni constant). Notice that in the final result we have restored the value $a_0$ of the scale factor evaluated at time $\tau_0$.

One may mow repeat the same steps to derive the tensor power spectrum. Here we limit ourselves to write the final result which is found to be:
\be
\mathcal P_{h} (k)  = \mathcal P_{h}^{(0)} + \Delta \mathcal P_h (k) , 
\ee
where:
\bea
\P_{h}^{(0)} &=&  \frac{2 H_0^2 }{ \pi^2 }  , \\
\frac{ \Delta \mathcal P_h (k) }{\P_{h}^{(0)} }&=& - 2 \epsilon_0   \ln ( k/ a_0 H_0) + 2 \epsilon_0  (\C - 1) .
\eea
Notice that in the present article we have introduced the label $h$ to label the horizon crossing time of tensor modes, which happens at a different time from the respective horizon crossing of scalars. This amounts to replace $0 \to h$ in every background quantity in our last expressions.

\end{appendix}

\end{document}